\documentstyle[epsf,graphicx,rotating]{mn}
\input{psfig}
\newdimen\digitwidth
\setbox0=\hbox{\rm0}
\digitwidth=\wd0
\catcode `@=\active
\def@{\kern\digitwidth}

\begin{document}
\def\la{\mathrel{\hbox{\rlap{\hbox{\lower4pt\hbox{$\sim$}}}\hbox{$<$}}}}
\def\ga{\mathrel{\hbox{\rlap{\hbox{\lower4pt\hbox{$\sim$}}}\hbox{$>$}}}}

\title[The classification of BL Lacertae objects]
{The classification of BL Lacertae objects: \\
the Ca H\&K break}

\author[H. Landt et al.]{Hermine Landt$^{1,2}$, Paolo Padovani$^{1,3,4}$, Paolo Giommi$^{5}$\\
$^1$ Space Telescope Science Institute, 3700 San Martin Drive, Baltimore, MD 21218, USA \\
$^2$ Hamburger Sternwarte, Gojenbergsweg 112, D-21029 Hamburg, Germany \\
$^3$ Affiliated to the Astrophysics Division, Space Science Department, European Space Agency \\
$^4$ On leave from Dipartimento di Fisica, II Universit\`a di Roma ``Tor Vergata'', Via della Ricerca 
Scientifica 1, I-00133 Roma, Italy \\
$^5$ ASI Science Data Center, c/o ESRIN, Via G. Galilei, I-00044 Frascati, Italy \\
}

\date{Accepted~~, Received~~}

\maketitle

\begin{abstract}
  
  We investigate why BL Lacertae objects (BL Lacs) have values of the
  Ca H\&K break (a stellar absorption feature) lower than low-power
  radio galaxies and if its use is justified to separate the two
  classes. For this purpose we relate this parameter to the radio and
  optical core emissions, as well as to the X-ray powers, for a sample
  of $\sim 90$ radio sources. We find that the Ca H\&K break value
  decreases with increasing jet powers, and that it also
  anti-correlates with the radio core dominance parameter but not with
  extended radio emission. Based on this we conclude that the Ca H\&K
  break value of BL Lacs and radio galaxies is a suitable indicator of
  orientation. From the luminosity ratios between objects with low and
  high Ca H\&K break values we constrain the average Lorentz factors
  for BL Lacs and low-power radio galaxies in the radio and X-ray band
  to $\Gamma \sim 2 - 4$ and derive average viewing angles for the
  galaxies. Our values are in agreement with results from independent
  methods. We find that the correlations between Ca H\&K break and
  radio core and X-ray luminosity hold independently for low- (LBL)
  and high-energy peaked BL Lacs (HBL). We derive average viewing
  angles for their parent populations, which turn out to be similar to
  the ones for our entire sample, and compare for the first time the
  luminosities of LBL and HBL at different orientations.

\end{abstract}


\section{Introduction}

BL Lacertae objects (BL Lacs) are set apart from other types of active
galactic nuclei by their extreme properties, namely, irregular, rapid
variability, strong optical and radio polarization, core-dominant
radio morphology, and a broad continuum extending from the radio
through the $\gamma$-rays. These properties are believed to be due to
the fact that BL Lacs are low-luminosity radio galaxies
(Fanaroff-Riley type I [FR I]; Fanaroff \& Riley 1974) whose jets are
aligned close to the observer's line of sight (first proposed by
Blandford \& Rees 1978). At such small viewing angles, relativistic
effects cause the jet's luminosity to appear amplified, and therefore
we refer to BL Lacs as ``beamed'' FR I radio galaxies.

The classification of an active galactic nucleus (AGN) as a BL Lac has
been controversial ever since the first surveys which included a
sizeable number of these objects, and has somewhat varied over time.
Two of the early surveys to select a large number of BL Lacs were the
1 Jy in the radio (Stickel et al. 1991; Rector \& Stocke 2001) and the
{\it EINSTEIN} Medium Sensitivity Survey (EMSS) in the X-ray band
(Stocke et al. 1991; Morris et al. 1991). The 1 Jy radio survey used
two criteria to classify an object as a BL Lac: a flat radio spectrum
($\alpha_{\rm r} \leq 0.5$, where $S_\nu \propto\nu^{-\alpha}$) and an
optical spectrum with emission lines weaker than 5~\AA~rest-frame
equivalent width. This latter criterion was chosen to separate BL Lacs
from their more powerful siblings, the flat spectrum radio quasars
(FSRQ). Other BL Lac properties, such as strong polarization and rapid
variability, were not regarded as necessary requirements.
Nevertheless, optical polarization studies, available at the time,
gave values above 3\% for almost all objects (K\"uhr \& Schmidt 1990),
and their strong and irregular variability was quantified later by
Heidt \& Wagner (1996).

The EMSS X-ray survey, on the other hand, did not use any radio
information to classify its objects (allowing for the existence of
radio-quiet X-ray emitting BL Lacs, which could not be found), but
introduced a limit on the Ca H\&K break value as an additional
criterion. The Ca H\&K break, also referred to as ``Ca II break'' or
``contrast'', is a stellar absorption feature typically found in the
spectra of elliptical galaxies. Its value in non-active elliptical
galaxies is on average $\sim 50\%$ (Dressler \& Shectman 1987). Stocke
et al. (1991) chose a limiting value of 25\% for the EMSS BL Lacs in
order to ensure the presence of a substantial non-thermal jet
continuum in addition to the thermal spectrum of the host galaxy. As
regards polarization (Stocke et al. 1991; see also Jannuzi et al.
1994) and variability (Heidt \& Wagner 1998) data, EMSS BL Lacs were
found to have somewhat lower values than the ones typical of 1 Jy BL
Lacs. However, similarly to the 1 Jy survey, these two BL Lac
properties were not used as part of the classification scheme.

The two defining criteria for a BL Lac used by the EMSS, a diluted Ca
H\&K break and no or very weak emission lines in the optical spectrum,
were revised by March\~a et al. (1996). In order to additionally
include in surveys radio sources with a strong host galaxy
contribution which might hide weak BL Lacs, these authors proposed to
expand the limit on the Ca H\&K break up to a value of 40\% and
accordingly, due to a lower continuum in these objects, the allowed
rest frame equivalent width of the strongest emission line up to a
value of $\sim$50~\AA. In fact, they suggested a triangular region in
the contrast, equivalent width plane (see their Fig. 6) to be used to
separate BL Lacs both from radio galaxies and quasars. BL Lacs
selected according to these extended criteria were termed ``BL Lac
candidates'', in order to distinguish them from those fulfilling the
classical criteria. Recent surveys, like RGB (Laurent-Muehleisen et
al. 1998), DXRBS (Perlman et al. 1998; Landt et al. 2001), the
Sedentary Survey (Giommi, Menna \& Padovani 1999), and REX (Caccianiga
et al. 1999, 2000), employ this revised version of the BL Lac
classification.

In this paper, we investigate why BL Lacs have lower Ca H\&K break
values than low-power radio galaxies. In Section 2, we relate the Ca
H\&K break value to optical jet power. In Section 3, we expand our
studies to the radio and X-ray band. In Section 4, we determine a
possible relation between Ca H\&K break value and viewing angle. An
application of our studies to the two BL Lac subclasses, low- and
high-energy peaked BL Lacs, is presented in Section 5. Finally, in
Section 6, we discuss our results and present our conclusions.
Throughout this paper, we assumed the following cosmological
parameters, $\rm H_0 = 50$ km/s/Mpc and $\rm q_0 = 0$.

\section{The Optical BL Lac Continuum}

According to current unified schemes for radio-loud AGN, BL Lacs are
beamed FR I radio galaxies. Therefore, in their optical spectra we see
the amplified non-thermal emission from the jet in addition to thermal
emission from the underlying host galaxy, normally a luminous
elliptical (e.g., Wurtz, Stocke \& Yee 1996). The non-thermal and
thermal components have different shapes, and their relative strengths
determine the observed BL Lac continuum. The jet component can be
locally best described by a single power law of the form $S_\nu
\propto \nu^{-\alpha_{\nu}}$ (or equivalently $S_\lambda \propto
\lambda^{-\alpha_{\lambda}}$, where $\alpha_{\lambda} = 2 -
\alpha_{\nu}$). The spectrum of the host galaxy, on the other hand,
has an approximate black body shape and contains absorption features
typical of ellipticals.

\setcounter{figure}{0}

\begin{figure}
\centerline{
\includegraphics[clip=true,bb=78 163 600 755,angle=-90,scale=0.45]
{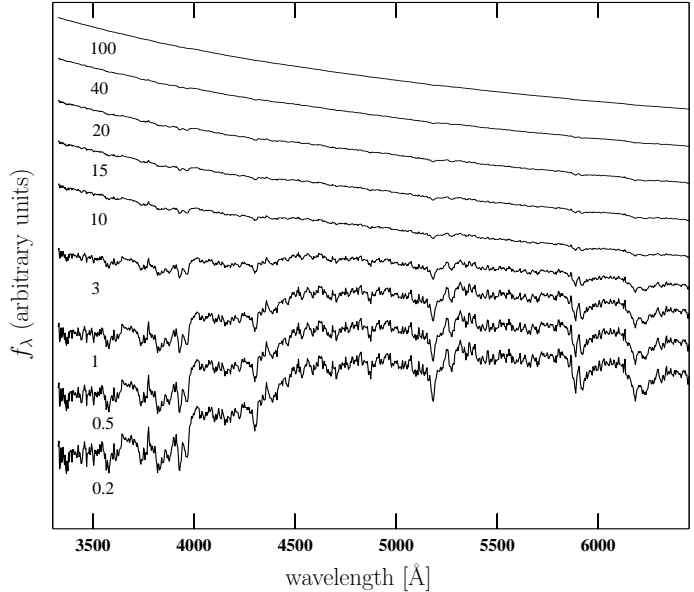}
}
\caption{Simulated BL Lac spectra $f_{\lambda}$ vs. $\lambda$ for a jet
  of optical spectral slope $\alpha_{\lambda}=\alpha_{\nu} = 1$ and of
  increasing flux relative to the underlying host galaxy. The assumed
  jet/galaxy ratios (defined at 5500 \AA) are from bottom to top: 0.2,
  0.5, 1, 3, 10, 15, 20, 40, and 100.}
\end{figure}

For our studies we have simulated possible BL Lac continua. For this
purpose, we assumed the jet overlaying the elliptical host galaxy to
vary in intensity and optical spectral slope. We used values between
0.2 and 100 for the jet/galaxy ratio (defined at 5500 \AA), and
assumed for the optical spectral slope the three cases $\alpha_\nu=0,
1, 2$, following the results of Falomo, Scarpa \& Bersanelli (1994),
and included additionally the extreme case of $\alpha_{\nu}=4$.

Fig. 1 shows the resulting simulated BL Lac spectra representatively
for a jet of optical spectral slope $\alpha_{\lambda}=\alpha_\nu=1$.
As the jet steadily increases relative to the galaxy, two effects are
visible. The shape of the BL Lac continuum resembles more and more a
power-law spectrum and the galactic absorption features become weaker.
One prominent absorption feature typically seen in the spectra of
elliptical galaxies is the Ca H\&K break located at $\sim$ 4000~\AA~
rest frame wavelength. This feature, also referred to as ``Ca II
break'' or ``contrast'', is defined as $C = (f_+ - f_-) / f_+$, where
$f_-$ and $f_+$ are the fluxes in the rest frame wavelength regions
$3750 - 3950$ \AA~ and $4050 - 4250$ \AA~ respectively. Its value in
normal non-active elliptical galaxies is found to be on average $\sim$
0.5 (Dressler \& Shectman 1987). In BL Lacs, the value of the Ca H\&K
break is decreased by the non-thermal jet emission. We note that the
Ca H\&K break value is related to the discontinuity at
4000~\AA~[D(4000); Bruzual 1983] by $C = 1 - 1/D(4000)$, and is
measured in spectra plotted as $f_{\nu}$ versus $\nu$. However, if
measured in spectra plotted as $f_{\lambda}$ versus $\lambda$, the
relation $C_{\nu} = 0.14 + 0.86 \cdot C_{\lambda}$ can be used to
convert one to the other.

In order to quantify how the amplified jet emission changes the Ca
H\&K break value, we measured this feature in each of our simulated
spectra. In the case where the jet dominates the object's spectrum, we
set the Ca H\&K break value equal to zero (its minimum possible value
in our simulations). For every case of $\alpha_\nu$, we get that there
is a significant ($P > 99.99\%$) linear anti-correlation between the
assumed jet/galaxy ratio and the Ca H\&K break value (dotted lines in
Fig. 2).

To assess how well this result is reproduced by observations we need
information on Ca H\&K break value and jet/galaxy ratios for a
sizeable sample of BL Lacs. Observational jet/galaxy ratios were
obtained from the imaging studies of Urry et al. (2000) which allow
the separation of the core flux from that of the host galaxy. These
authors used the HST Wide Field Planetary Camera 2 (WFPC 2) to image
in snapshot mode a sample of 132 BL Lacs from different radio and
X-ray surveys. Most of their images were taken in the F702W filter
(similar to Cousin R filter), but those already observed in F814W
during an earlier HST cycle were reobserved in F606W (a filter similar
to Johnson V). We derived jet/galaxy ratios at 5500 \AA~by converting
$R$ magnitudes to $V$ magnitudes. We assumed $V-R = 0.3$ for the core
component (Urry et al. 2000), while for the host galaxy magnitudes we
used the redshift-dependent $V-R$ values as tabulated by Urry et al.
(2000; see their Table 2). Ca H\&K break values for 48 of these
sources were derived from the literature and also from our own
measurements (see Sect. 3).

For these BL Lacs, we plot in Fig. 2 the jet/galaxy ratio vs. Ca H\&K
break value. An analysis using the ASURV package (Feigelson \& Nelson
1985), that allows us to include upper limits, shows that these two
quantities are significantly ($P > 99.99\%$) anti-correlated, in
agreement with our simulations. Furthermore, a partial correlation
analysis shows that this correlation is not induced by a common
redshift dependence. In Fig. 2 we plot the observed correlation
between optical jet/galaxy ratio and Ca H\&K break value (solid line)
in addition to the correlations obtained from our simulations (dotted
lines). A comparison between the two indicates that the scatter in
the observed correlation ($\sim 0.6$) is likely induced by jets of
different optical spectral slopes.

\begin{figure}
\centerline{\psfig
{figure=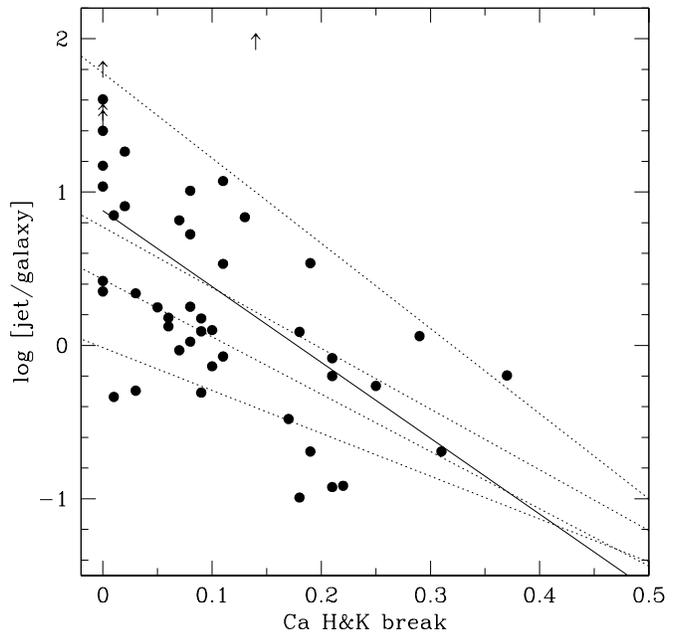,width=9cm}}
\caption{The jet/galaxy flux ratio (at 5500 \AA) vs. the Ca H\&K break
  value for BL Lacs from Urry et al. (2000). Arrows indicate lower
  limits on the jet/galaxy ratio. The solid line represents the
  observed correlation. Dotted lines represent the correlations
  obtained from our simulations for optical spectral slopes
  $\alpha_{\nu} = 0,1,2,4$ (from bottom to top).}
\end{figure}

\begin{figure}
\centerline{\psfig
{figure=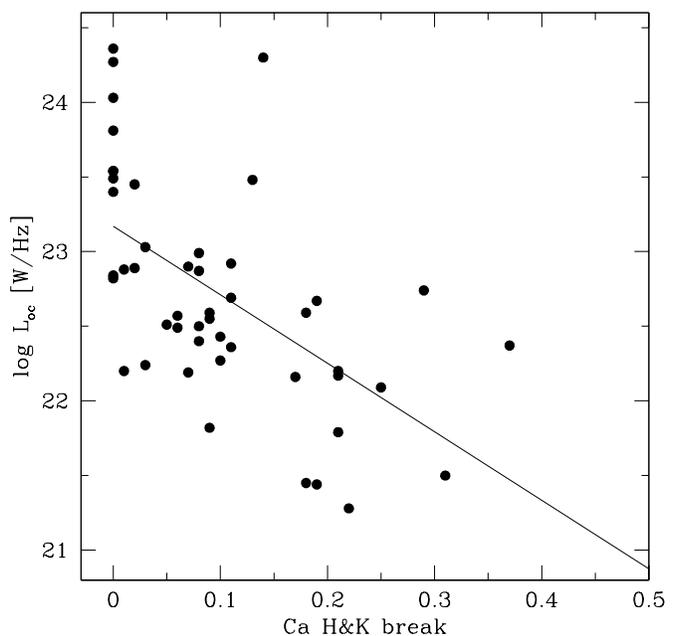,width=9cm}}
\caption{The optical core luminosity vs. the Ca H\&K break value for BL
  Lacs from Urry et al. (2000). The solid line represents the observed
  correlation.}
\end{figure}

We have shown using both simulations and observational data that the
Ca H\&K break value in BL Lacs anti-correlates with optical jet/galaxy
ratio. At this point the question arises if this correlation could
possibly mask a correlation between jet emission and Ca H\&K break
value or between host galaxy emission and Ca H\&K break value. Urry et
al. (2000) find that the distribution of the absolute magnitudes of BL
Lac host galaxies has a relatively small dispersion $\Delta M_R = 0.6$
mag. This result is based on detected host galaxies of a very large
number of BL Lacs (72 sources) covering the redshift range $0.024 \le
z \le 0.7$. The relatively small dispersion found by Urry et al.
(2000) implies that the luminosity of BL Lac host galaxies can be
regarded as roughly constant. Therefore, the observed increase in
jet/galaxy ratio (with decreasing Ca H\&K break) can be related
directly to an increase in jet power. Indeed, for our sample of
objects we find that, on one hand, the optical core luminosity (from
Urry et al. 2000) is significantly ($P>99.9\%$) anti-correlated with
the Ca H\&K break value (see Fig.  3), and that, on the other hand, no
significant ($P=88\%$) correlation is present between host galaxy
luminosity and Ca H\&K break value.

\begin{table*}
\begin{minipage}{180mm}
{\bf Table 1.} BL Lacs and FR I radio galaxies \\[0.1cm] \nobreak
\begin{tabular}{lllrccrccccrrr} 
\hline 
Survey & Name & \multicolumn{1}{c}{z} & \multicolumn{1}{c}{C} & $\sigma$ & Ref. & 
\multicolumn{1}{c}{log ${\rm L_{\rm x}}$} & Ref. & log ${\rm L_{\rm rc}}$ & 
Ref. & \multicolumn{1}{c}{Radio} & \multicolumn{1}{c}{log ${\rm L_{\rm ext}}$} & \multicolumn{1}{c}{log R} & 
\multicolumn{1}{c}{$\alpha_{\rm r}$} \\
&&&&& C & 1 keV & $\rm f_{\rm x}$ & 5 GHz & $\rm f_{\rm rc}$ & \multicolumn{1}{c}{Array} &
\multicolumn{1}{c}{5 GHz} & \multicolumn{1}{c}{5 GHz} & \\
&&&&&& [W/Hz] && [W/Hz] &&& \multicolumn{1}{c}{[W/Hz]} && \\
(1) & (2) & \multicolumn{1}{c}{(3)} & \multicolumn{1}{c}{(4)} & (5) & (6) & 
\multicolumn{1}{c}{(7)} & (8) & (9) & (10) & \multicolumn{1}{c}{(11)} & \multicolumn{1}{c}{(12)} & 
\multicolumn{1}{c}{(13)} & \multicolumn{1}{c}{(14)} \\
\hline
1 Jy    & 0118$-$272      & 0.559 & 0.00 & 0.02 & v & 20.45 & a & 26.94 & m & IV   & 26.81 & $ $0.13 & $-$0.16 \\
1 Jy    & 0138$-$097      & 0.733 & 0.00 & 0.03 & v & 20.56 & a & 27.12 & m & IV   & 26.99 & $ $0.13 & $-$0.50 \\
1 Jy    & 0218$+$357      & 0.685 & 0.12 & 0.09 & v &$<20.51$&A & 27.25 & o & XI   & 27.07 & $ $0.18 & $ $0.31 \\
1 Jy    & 0828$+$493      & 0.548 & 0.08 & 0.06 & v &$<20.29$&A & 26.55 & y & II   & 26.78 & $-$0.23 & $-$0.88 \\  
1 Jy    & 1538$+$149      & 0.605 & 0.00 & 0.03 & v & 20.77 & x & 27.13 & y & II   & 27.19 & $-$0.06 & $-$0.28 \\
1 Jy    & 1823$+$568      & 0.664 & 0.02 & 0.03 & v & 21.02 & x & 27.05 & y & II   & 27.31 & $-$0.26 & $-$0.13 \\   
2 Jy    & 0625$-$53       & 0.054 & 0.15 & 0.12 & v & 19.32 & b & 23.75 & p & IX   & 25.37 & $-$1.62 & $ $1.06 \\
2 Jy    & 0915$-$11       & 0.054 & 0.36 & 0.01 & w & 19.84 & b & 24.45 & p & VII  & 26.24 & $-$1.79 & $ $0.82 \\
200 mJy & 0055$+$300      & 0.015 & 0.46 &      & u &$<16.87$&A & 23.22 & n & X    & 22.93 & $ $0.29 & $ $0.90 \\
200 mJy & 0116$+$319      & 0.060 & 0.45 &      & u &$<18.09$&A & 25.09 & n & X    & 25.11 & $-$0.02 & $ $0.42 \\
200 mJy & 0149$+$710      & 0.022 & 0.33 &      & u & 17.86 & c & 23.78 & q & III  & 23.87 & $-$0.09 & $-$0.10 \\
200 mJy & 0210$+$515      & 0.049 & 0.15 &      & u & 19.19 & c & 24.22 & q & III  & 24.09 & $ $0.13 & $ $0.04 \\
200 mJy & 0651$+$410      & 0.021 & 0.39 &      & u &$<17.19$&A & 23.74 & o & IV   & 23.39 & $ $0.35 & $-$0.51 \\
200 mJy & 0651$+$428      & 0.126 & 0.17 &      & u & 18.93 & c & 24.97 & q & III  & 24.55 & $ $0.42 & $ $0.06 \\
200 mJy & 0733$+$597      & 0.041 & 0.49 &      & u &$<17.88$&A & 24.23 & q & III  & 23.97 & $ $0.26 & $ $0.36 \\
200 mJy & 1055$+$567      & 0.410 & $\le0.05$ & & u & 20.77 & c & 26.11 & q & III  & 25.76 & $ $0.35 & $-$0.10 \\
200 mJy & 1101$+$384      & 0.031 & 0.24 &      & u & 20.28 & d & 24.40 & q & VIII & 23.71 & $ $0.69 & $ $0.05 \\
200 mJy & 1133$+$704      & 0.046 & 0.31 &      & u & 19.79 & d & 24.11 & q & III  & 24.09 & $ $0.02 & $ $0.15 \\
200 mJy & 1144$+$352      & 0.063 & 0.37 &      & u & 18.07 & d & 24.75 & q & III  & 24.75 & $ $0.00 & $-$0.03 \\
200 mJy & 1217$+$295      & 0.002 & 0.42 &      & u & 17.76 & A & 21.69 & q & VIII & 20.86 & $ $0.83 & $ $0.13 \\
200 mJy & 1241$+$735      & 0.075 & 0.43 &      & u &$<15.00$&A & 24.52 & n & X    & 24.70 & $-$0.18 & $-$0.13 \\
200 mJy & 1418$+$546      & 0.151 & 0.03 &      & u & 19.20 & d & 26.10 & q & VIII & 25.46 & $ $0.64 & $-$0.62 \\
200 mJy & 1652$+$398      & 0.031 & 0.07 &      & u & 19.59 & d & 24.79 & m & IV   & 23.46 & $ $1.33 & $ $0.10 \\
200 mJy & 1744$+$260      & 0.147 &$\le0.3$&    & u &$<18.74$&A & 25.11 & n & X    & 25.09 & $ $0.02 & $ $0.23 \\ 
200 mJy & 1807$+$698      & 0.046 & 0.03 &      & u & 18.43 & d & 25.14 & m & IV   & 24.79 & $ $0.35 & $-$0.12 \\
200 mJy & 2320$+$203      & 0.038 & 0.47 &      & u & 17.63 & d & 23.85 & q & III  & 24.18 & $-$0.33 & $ $0.03 \\
EMSS    & MS0011.7$+$0837 & 0.162 & 0.30 & 0.04 & e & 19.01 & e & 24.64 & e & II   & 24.80 & $-$0.16 & $ $1.13 \\
EMSS    & MS0122.1$+$0903 & 0.338 & 0.22 & 0.18 & s & 19.18 & f & 23.88 & s & II   &$<23.74$&$>0.14$ &         \\
EMSS    & MS0158.5$+$0019 & 0.298 & 0.09 & 0.02 & s & 20.56 & f & 24.37 & z & II   & 24.35 & $ $0.02 & $ $0.13 \\
EMSS    & MS0257.9$+$3429 & 0.246 & 0.25 & 0.06 & s & 19.13 & f & 24.34 & z & II   & 23.70 & $ $0.64 & $ $0.10 \\
EMSS    & MS0317.0$+$1834 & 0.190 & 0.21 & 0.06 & s & 19.57 & f & 24.05 & z & II   & 24.21 & $-$0.16 & $ $0.24 \\ 
EMSS    & MS0419.3$+$1943 & 0.516 & 0.11 & 0.07 & s & 20.64 & f & 24.92 & s & II   & 24.23 & $ $0.69 & $ $0.09 \\
EMSS    & MS0607.9$+$7108 & 0.267 & 0.09 & 0.04 & s & 19.04 & f & 24.52 & z & II   & 24.43 & $ $0.09 & $ $0.31 \\
EMSS    & MS0737.9$+$7441 & 0.314 & 0.00 & 0.01 & s & 20.51 & f & 24.86 & z & II   & 24.48 & $ $0.38 & $-$0.02 \\
EMSS    & MS0922.9$+$7459 & 0.638 & 0.20 & 0.07 & s & 20.57 & f & 24.88 & s & II   &$<24.03$&$>0.85$ &         \\
EMSS    & MS1019.0$+$5139 & 0.141 & 0.23 & 0.05 & s & 19.73 & e & 23.31 & s & V    & 22.29 & $ $1.02 & $ $0.61 \\
EMSS    & MS1050.7$+$4946 & 0.140 & 0.32 & 0.07 & s & 19.46 & e & 24.50 & s & V    & 24.18 & $ $0.32 & $ $0.15 \\
EMSS    & MS1154.1$+$4255 & 0.172 & 0.33 & 0.10 & s & 18.95 & e & 24.08 & e & V    & 23.17 & $ $0.91 & $ $0.57 \\
EMSS    & MS1207.9$+$3945 & 0.616 & 0.07 & 0.04 & s & 20.90 & f & 25.53 & s & V    & 24.59 & $ $0.94 & $ $0.95 \\
EMSS    & MS1209.0$+$3917 & 0.602 & 0.15 & 0.13 & s & 20.27 & e & 25.28 & e & V    &$<24.31$&$>0.97$ & $ $0.67 \\
EMSS    & MS1221.8$+$2452 & 0.218 & 0.02 & 0.02 & s & 19.51 & f & 24.60 & z & II   & 24.14 & $ $0.46 & $-$0.02 \\
EMSS    & MS1229.2$+$6430 & 0.164 & 0.18 & 0.09 & s & 20.16 & f & 24.56 & z & II   & 24.18 & $ $0.38 & $ $0.27 \\
EMSS    & MS1235.4$+$6315 & 0.297 & 0.05 & 0.07 & s & 19.87 & f & 24.62 & s & V    &$<23.24$&$>1.38$ & $ $0.47 \\
EMSS    & MS1407.9$+$5954 & 0.496 & 0.10 & 0.05 & s & 20.12 & f & 25.15 & z & II   & 24.97 & $ $0.18 & $ $0.64 \\
EMSS    & MS1443.5$+$6349 & 0.298 & 0.22 & 0.06 & s & 19.98 & f & 24.40 & z & II   & 24.40 & $ $0.00 & $ $0.39 \\
EMSS    & MS1458.8$+$2249 & 0.235 & 0.00 & 0.01 & s & 20.14 & f & 24.69 & z & II   & 24.37 & $ $0.32 & $ $0.07 \\
EMSS    & MS1534.2$+$0148 & 0.311 & 0.11 & 0.06 & s & 20.29 & f & 25.00 & z & II   & 24.84 & $ $0.16 & $ $0.61 \\ 
EMSS    & MS1552.1$+$2020 & 0.273 & 0.13 & 0.09 & s & 20.71 & f & 24.94 & z & II   & 24.73 & $ $0.21 & $ $0.60 \\
EMSS    & MS1757.7$+$7034 & 0.406 & 0.01 & 0.00 & s & 20.68 & f & 24.78 & s & II   &$<24.27$&$>0.51$ &         \\
EMSS    & MS2143.4$+$0704 & 0.235 & 0.10 &      & j & 19.94 & f & 24.93 & z & II   & 24.70 & $ $0.23 & $ $0.57 \\
EMSS    & MS2347.4$+$1924 & 0.515 & 0.18 & 0.16 & s & 20.28 & s & 24.57 & s & II   & 23.88 & $ $0.69 & $ $0.36 \\
SLEW    & 1ES0120$+$340   & 0.272 & 0.00 & 0.04 & v & 21.41 & h & 25.03 & r & VI   & 24.01 & $ $1.02 & $ $0.24 \\ 
SLEW    & 1ES0229$+$200   & 0.139 & 0.21 & 0.07 & v & 20.34 & g & 24.60 & r & VI   & 22.97 & $ $1.63 & $ $0.48 \\
SLEW    & 1ES0502$+$675   & 0.416$^\ast$ & 0.00 & 0.04 & v & 21.70 & h & 25.21 & q & I & 24.52 & $ $0.69 & $ $0.04\\ 
SLEW    & 1ES0806$+$524   & 0.138 & 0.00 & 0.04 & v & 20.31 & h & 25.01 & q & III  & 24.63 & $ $0.38 & $ $0.03 \\
SLEW    & 1ES0927$+$500   & 0.188 & 0.03 & 0.07 & v & 20.73 & i & 24.36 & q & III  & 24.04 & $ $0.32 & $-$0.02 \\
SLEW    & 1ES1255$+$244   & 0.140 & 0.08 & 0.11 & v & 20.92 & r & 23.80 & r & VI   & 22.65 & $ $1.15 & $ $0.55 \\ 
SLEW    & 1ES1426$+$428   & 0.129 & 0.01 &      & j & 20.75 & h & 24.20 & q & I    & 24.09 & $ $0.32 & $ $0.35 \\
SLEW    & 1ES1440$+$122   & 0.162 & 0.18 &      & j & 20.22 & h & 24.73 & q & VIII & 23.71 & $ $1.02 & $ $0.26 \\
\hline
\end{tabular}
\end{minipage}
\end{table*}

\begin{table*}
\begin{minipage}{180mm}
\setcounter{table}{1}
\contcaption{}
\begin{tabular}{lllrccrccccrrr} 
\hline 
Survey & Name & \multicolumn{1}{c}{z} & \multicolumn{1}{c}{C} & $\sigma$ & Ref. & 
\multicolumn{1}{c}{log ${\rm L_{\rm x}}$} & Ref. & log ${\rm L_{\rm rc}}$ & 
Ref. & \multicolumn{1}{c}{Radio} & \multicolumn{1}{c}{log ${\rm L_{\rm ext}}$} & \multicolumn{1}{c}{log R} & 
\multicolumn{1}{c}{$\alpha_{\rm r}$} \\
&&&&& C & 1 keV & $\rm f_{\rm x}$ & 5 GHz & $\rm f_{\rm rc}$ & \multicolumn{1}{c}{Array} &
\multicolumn{1}{c}{5 GHz} & \multicolumn{1}{c}{5 GHz} & \\
&&&&&& [W/Hz] && [W/Hz] &&& \multicolumn{1}{c}{[W/Hz]} && \\
(1) & (2) & \multicolumn{1}{c}{(3)} & \multicolumn{1}{c}{(4)} & (5) & (6) & 
\multicolumn{1}{c}{(7)} & (8) & (9) & (10) & \multicolumn{1}{c}{(11)} & \multicolumn{1}{c}{(12)} & 
\multicolumn{1}{c}{(13)} & \multicolumn{1}{c}{(14)} \\
\hline
SLEW    & 1ES1741$+$196     & 0.083 & 0.12 & 0.05 & v & 19.55 & h & 24.67 & q & III  & 24.73 & $-$0.06 & $-$0.07 \\
SLEW    & 1ES1853$+$671     & 0.212 & 0.09 & 0.05 & v & 20.20 & r & 24.37 & r & VI   &$<22.87$&$>1.50$ & $-$0.07 \\ 
SLEW    & 1ES2326$+$174     & 0.213 & 0.06 & 0.05 & v & 20.50 & r & 24.55 & r & VI   & 24.23 & $ $0.32 & $-$0.03 \\  
RGB     & RGBJ0110$+$418    & 0.096 & 0.32 &      & j & 19.08 & j & 23.88 & q & I    & 23.88 & $ $0.00 & $ $0.70 \\
RGB     & RGBJ0152$+$017    & 0.080 & 0.29 &      & j & 19.22 & j & 24.26 & q & VIII &       & $ $     & $ $0.12 \\
RGB     & RGBJ0314$+$247    & 0.054 & 0.30 &      & j & 18.50 & j & 22.88 & q & VIII & 23.65 & $-$0.77 & $-$1.16 \\
RGB     & RGBJ0656$+$426    & 0.059 & 0.36 &      & j & 18.84 & j & 24.33 & q & III  & 24.72 & $-$0.39 & $ $0.53 \\
RGB     & RGBJ0710$+$591    & 0.125 & 0.22 &      & j & 20.49 & j & 24.39 & q & I    & 24.54 & $-$0.15 & $ $0.54 \\
RGB     & RGBJ0806$+$728    & 0.098 & 0.16 &      & j & 19.23 & j & 23.93 & q & I    & 23.67 & $ $0.26 & $ $0.39 \\
RGB     & RGBJ0820$+$488    & 0.130 & 0.41 &      & j & 18.96 & j & 23.59 & q & I    & 24.84 & $-$1.25 & $ $0.59 \\
RGB     & RGBJ1136$+$676    & 0.136 & 0.20 &      & j & 20.45 & j & 24.50 & q & I    & 23.81 & $ $0.69 & $-$0.04 \\
RGB     & RGBJ1253$+$509    & 0.121 & 0.49 &      & j & 18.93 & j & 23.67 & q & I    & 24.33 & $-$0.66 & $ $0.51 \\
RGB     & RGBJ1324$+$576    & 0.115 & 0.41 &      & j & 19.36 & j & 24.16 & q & I    & 23.98 & $ $0.18 & $ $0.04 \\
RGB     & RGBJ1417$+$257    & 0.237 & 0.00 &      & j & 20.95 & j & 25.05 & q & VIII &       & $ $     & $ $0.67 \\
RGB     & RGBJ1427$+$541    & 0.105 & 0.39 &      & j & 18.68 & j & 24.07 & q & VI   & 23.65 & $ $0.42 & $ $0.24 \\
RGB     & RGBJ1516$+$293    & 0.130 & 0.32 &      & j & 19.30 & j & 24.41 & q & I    &       & $ $     &         \\
RGB     & RGBJ1532$+$302    & 0.064 & 0.29 &      & j & 19.09 & j & 23.92 & q & III  & 23.16 & $ $0.76 & $-$0.01 \\
RGB     & RGBJ1745$+$398    & 0.267 & 0.17 &      & j & 19.86 & j & 25.64 & q & III  & 25.70 & $-$0.06 & $ $0.75 \\
RGB     & RGBJ1750$+$470    & 0.160 & 0.29 &      & j & 19.71 & j & 24.08 & q & I    & 24.66 & $-$0.58 & $ $0.59 \\
RGB     & RGBJ1823$+$334    & 0.108 & 0.49 &      & j & 19.27 & j & 23.22 & q & I    & 24.50 & $-$1.28 & $ $0.91 \\
RGB     & RGBJ1841$+$591    & 0.530 & 0.17 &      & j & 20.24 & j & 24.94 & q & I    &       & $ $     &         \\
RGB     & RGBJ2241$+$048    & 0.069 & 0.45 &      & j & 18.36 & j & 23.94 & q & III  & 24.14 & $-$0.20 & $ $0.42 \\
RGB     & RGBJ2250$+$384    & 0.119 & 0.07 &      & j & 19.76 & j & 24.56 & q & III  & 24.56 & $ $0.00 & $-$0.11 \\  
RGB     & RGBJ2322$+$346    & 0.098 & 0.33 &      & j & 19.05 & j & 24.10 & q & I    & 24.30 & $-$0.20 & $ $0.17 \\
RGB     & RGBJ2323$+$205    & 0.041 & 0.48 &      & j & 17.70 & j & 23.92 & q & III  & 24.25 & $-$0.33 & $ $0.03 \\
DXRBS   & WGAJ0032.5$-$2849 & 0.324 & 0.22 & 0.08 & k & 19.11 & k & 25.79 & t & IX   & 24.88 & $ $0.91 & $ $0.08 \\ 
DXRBS   & WGAJ0340.8$-$1814 & 0.195 & 0.40 & 0.08 & k & 17.85 & k & 24.53 & t & IX   & 25.37 & $-$0.84 & $ $0.57 \\ 
DXRBS   & WGAJ0428.8$-$3805 & 0.150 & 0.32 & 0.05 & k & 17.99 & k & 24.50 & t & IX   & 24.24 & $ $0.26 & $-$0.03 \\ 
DXRBS   & WGAJ0528.5$-$5820 & 0.254 & 0.38 & 0.18 & l & 18.76 & k & 25.01 & t & IX   & 25.29 & $-$0.28 & $ $0.46 \\ 
DXRBS   & WGAJ0624.7$-$3230 & 0.252 & 0.22 & 0.05 & k & 19.26 & k & 24.72 & t & IX   & 25.29 & $-$0.47 & $-$0.54 \\ 
DXRBS   & WGAJ0816.0$-$0736 & 0.040 & 0.37 & 0.18 & k & 16.98 & k & 23.41 & t & IX   & 23.20 & $ $0.21 & $-$0.28 \\ 
DXRBS   & WGAJ1057.6$-$7724 & 0.181 & 0.48 & 0.18 & l & 18.14 & k & 25.38 & t & IX   & 25.66 & $-$0.28 & $ $0.71 \\ 
DXRBS   & WGAJ1311.3$-$0521 & 0.160 & 0.33 & 0.12 & l & 18.61 & l & 24.53 & t & IX   & 24.30 & $ $0.23 & $ $0.37 \\ 
DXRBS   & WGAJ1457.9$-$2124 & 0.319 & 0.45 & 0.20 & l & 19.57 & l & 25.25 & t & IX   & 25.80 & $-$0.55 & $ $0.79 \\ 
DXRBS   & WGAJ2317.4$-$4213 & 0.056 & 0.52 & 0.08 & k & 16.58 & k & 23.28 & t & IX   & 24.22 & $-$0.94 & $ $0.53 \\
\hline
\end{tabular}

columns: (1) survey, (2) object name, (3) redshift, (4) Ca H\&K break
value (measured in spectra $f_{\lambda}$ vs. $\lambda$), (5) $1\sigma$
error on the Ca H\&K break value, (6) reference for Ca H\&K break
measurement, (7) X-ray luminosity at 1 keV (ROSAT X-ray band (0.1 -
2.4 keV) flux transformed to 1 keV and luminosity k-corrected assuming
spectral index $\alpha_{\rm x} = 1.2$), (8) reference for ROSAT X-ray
band flux, (9) radio core luminosity at 5 GHz (transformed to 5 GHz
and k-corrected assuming $\alpha_{\rm r}$ from column (14) or else
$\alpha_{\rm r} = 0.3$), (10) reference for radio core flux, (11)
radio array used for observations of radio core flux (see below), (12)
extended radio luminosity at 5 GHz, (13) radio core dominance, (14)
radio spectral index between 1.4 (NVSS) and 5 GHz (computed from the
object's total fluxes)

\vspace*{0.1cm}

radio array: (I) VLA A at 5 GHz, (II) VLA A at 1.5 GHz, (III) VLA BnA
at 5 GHz, (IV) VLA B at 5 GHz, (V) VLA B at 1.5 GHz, (VI) VLA CnB at 5
GHz, (VII) VLA C at 5 GHz, (VIII) VLA D at 5 GHz, (IX) ATCA at 5 GHz,
(X) MERLIN at 5 GHz, (XI) VLBI at 1.5 GHz

\vspace*{0.1cm}

references: (a) Siebert et al. 1998, (b) Brinkmann, Siebert \& Boller
1994, (c) Brinkmann et al. 1997, (d) Brinkmann et al. 1995, (e)
Rector, Stocke \& Perlman 1999, (f) Perlman et al. 1996a, (g) Nass et
al. 1996, (h) Laurent-Muehleisen et al. 1999, (i) Bade et al. 1998,
(j) Laurent-Muehleisen et al. 1998, (k) Perlman et al. 1998, (l) Landt
et al. 2001, (m) Cassaro et al. 1999, (n) Augusto, Wilkinson \&
Browne 1998, (o) Taylor et al. 1996, (p) Morganti, Killeen \& Tadhunter
1993, (q) Laurent-Muehleisen et al. 1997, (r) Perlman et al. 1996b,
(s) Rector et al. 2000, (t) from ATCA observations of the DXRBS, (u)
March\~a et al. 1996, (v) measured on spectrum, (w) Owen, Ledlow \&
Keel 1996, (x) RASS/NVSS-ASDC catalogue (Giommi et al. in prep.), (y)
Murphy, Browne \& Perley 1993, (z) Perlman \& Stocke 1993, (A) ROSAT
All Sky Survey (Voges et al. 1999); upper limit estimated from
exposure maps

\vspace*{0.1cm}

$^\ast$ we obtained this redshift, which is different from the one
published by Perlman et al. (1996b), after a careful reinspection of
the electronic spectral file

\end{minipage}
\end{table*}

\section{From a BL Lac to a Radio Galaxy}

\begin{figure}
\centerline{\psfig
{figure=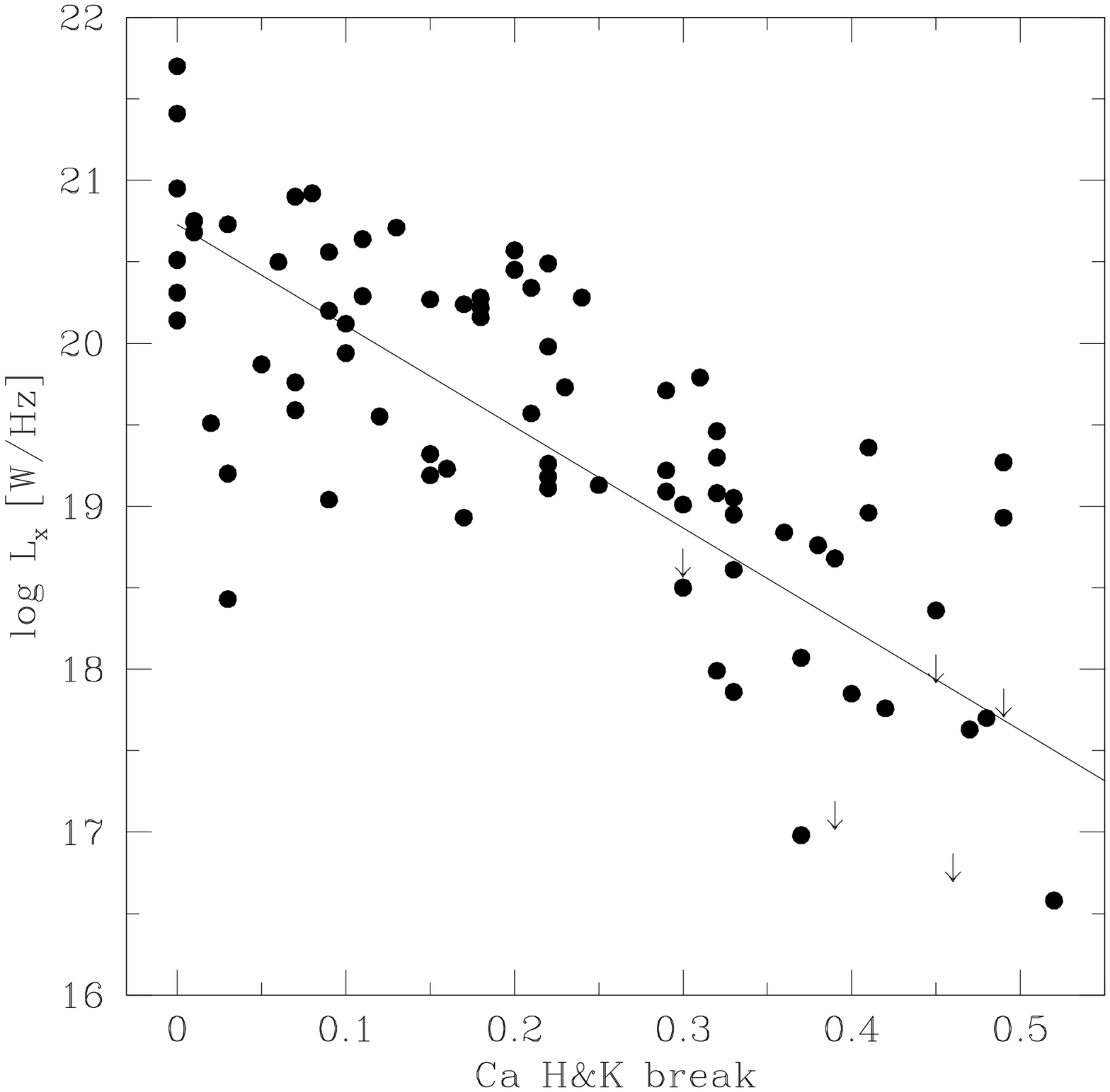,width=9cm}}
\caption{The 1 keV X-ray luminosity vs. the Ca H\&K break value. Arrows
  denote upper limits. The solid line represents the observed
  correlation. Object 1241$+$735 is off the plot.}
\end{figure}

\begin{figure}
\centerline{\psfig
{figure=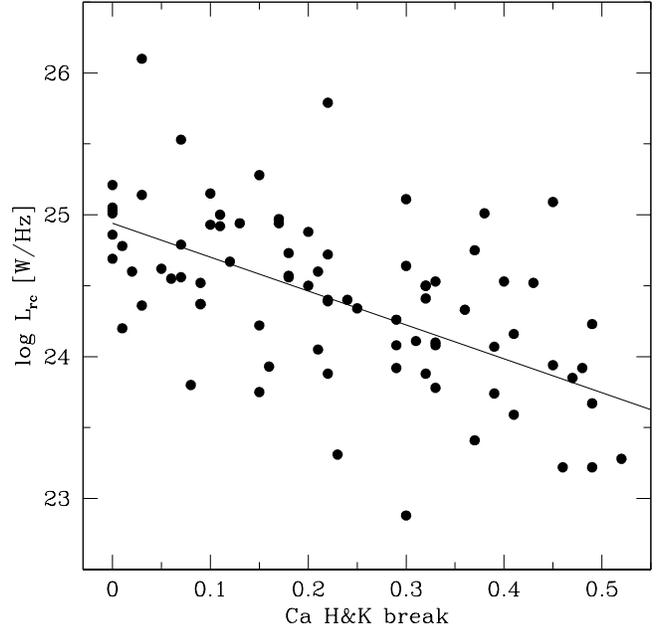,width=9cm}}
\caption{The radio core luminosity at 5 GHz vs. the Ca H\&K break
  value. The solid line represents the observed correlation.}
\end{figure}

In the previous section, we have shown that the Ca H\&K break value in
BL Lacs is decreased due to an increase in optical jet power. Here we
want to investigate this further by relating the Ca H\&K break value
also to the radio and X-ray properties of both BL Lacs and FR I radio
galaxies.

We have then collected from the literature and from our own data
information on the Ca H\&K break value, ROSAT X-ray band ($0.1 - 2.4$
keV) flux, total radio flux at 1.4 (NVSS) and 5 GHz, and core radio
flux. From these we have computed the luminosities, radio core
dominance parameter and radio spectral index $\alpha_{\rm r}$ which
are listed in Table 1 (see footnotes to Table 1 for details). We also
give in Table 1 the references for the Ca H\&K break value, and the
radio core and X-ray fluxes. The radio core flux measurements are
largely based on observations with the Very Large Array (VLA). The
total radio flux at 5 GHz was obtained from the GB6 survey (Gregory et
al. 1996), with the following exceptions: for the 1 and 2 Jy objects
we used the values given in Stickel et al. (1994) and Wall \& Peacock
(1985), for the EMSS objects we used the measurements given in Stocke
et al. (1991), for Slew objects not detected in the GB6 we used the
values given in Perlman et al. (1996b), and for southern DXRBS sources
we give the total radio flux at 5 GHz from the PMN survey (Griffith \&
Wright 1993).

The BL Lacs and FR I radio galaxies were selected from the following
surveys: 1 Jy survey (Stickel, Meisenheimer \& K\"uhr 1994), 2 Jy
survey (Tadhunter et al. 1993), 200 mJy sample (March\~a et al. 1996),
EMSS (Rector, Stocke \& Perlman 1999, 2000), Slew survey (Perlman et
al. 1996b), RGB (Laurent-Muehleisen et al. 1998), and DXRBS (Perlman
et al. 1998; Landt et al. 2001). From these surveys we chose in fact
all objects with a rest frame equivalent width of the strongest
emission line $<60$~\AA, an available Ca H\&K break value, and {\it
  both} a radio core and X-ray flux measurement. We required available
fluxes in both radio and X-ray band in order to avoid comparing two
different sets of objects in the luminosity - Ca H\&K break plane (see
Sect. 3.1). However, while for a particular object it is always
possible to obtain an upper limit on its ROSAT X-ray luminosity from
the exposure maps of the RASS (Voges et al. 1999), this is not the
case for its radio core luminosity and the object will not be included
in our sample. The limit on the emission line strength is based on the
results of March\~a et al. (1996, see their Fig. 6) and was chosen to
exclude flat spectrum radio quasars. An exclusion of objects whose
power is typical of FR II radio galaxies can be based on the value of
the extended radio emission. Following the results of Owen \& Ledlow
(1994), we chose a value of $L_{\rm ext} = 10^{25.6}$ W/Hz
(transformed from 1400 MHz using a radio spectral index $\alpha_{\rm
  r} = 0.8$) to separate the less luminous FR I radio galaxies from
the more luminous FR II radio galaxies. These authors find that
basically only FR II radio galaxies have an extended radio emission
above this value. The extended radio power was computed as $L_{\rm
  ext} = L_{\rm tot} - L_{\rm rc}$, where $L_{\rm tot}$ and $L_{\rm
  rc}$ are the object's total and core radio powers at 5 GHz
respectively. We note that we {\it assume} that our low-power radio
galaxies are FR Is, although we do not have any information on their
radio morphology. Bondi et al. (2001) have indeed shown that the
nuclear properties of the weak-lined radio galaxies in the 200 mJy
sample are consistent with those of FR Is, although their radio
morphology might not be consistent in all cases with an FR I
classification. Our sample comprises 83 objects. However, in Table 1
we additionally list 7 objects that meet the current classification
criteria for a BL Lac but have extended radio powers typical of FR II
radio galaxies. We will use these sources only for our comparison
studies in Sect. 5.2.

We measured the Ca H\&K break values for objects from the 1 Jy, 2 Jy,
Slew survey, and DXRBS on the reduced and calibrated spectra. For
sources from the 200 mJy sample, EMSS, and RGB, we used the Ca H\&K
break values listed in the literature. We note that we excluded from
our study sources with errors on the Ca H\&K break value $> 0.2$. The
error on the Ca H\&K break listed in Table 1 represents the 1 $\sigma$
limit and was computed based on the signal-to-noise ratio (SNR)
blueward and redward of the feature. The error is on average 0.07 for
the sources included in our study. No errors were available for
sources from the 200 mJy sample and RGB. Nevertheless, the spectra of
the RGB objects are quoted to have a SNR $\ga 30$ (Laurent-Muehleisen
et al. 1998) and therefore the error on the Ca H\&K break value for
these sources is expected to be small.

Our sample is highly heterogeneous, being a collection of surveys with
different selection bands and flux limits. However, similarly to the
studies of Padovani \& Giommi (1996), this approach offers us the
possibility to maximize the coverage of the parameter space and
therefore to look for correlations in a way which would not be
possible by considering individual samples separately. This approach
is acceptable whenever the parameter values or correlations studied
are not strongly influenced by the sample selection.

\subsection{The radio core and X-ray luminosity}

In the optical band, the spectrum of a BL Lac is made up of two
components, a thermal (galaxy) and a non-thermal (jet) one. But this
is not the case for the radio and X-ray bands, where we observe mainly
the jet emission. (Though in the X-ray case extended emission from
cluster- or group-scale gas can contribute to the observed
luminosity.) Therefore, following our results from Section 2, we
expect the radio core and X-ray luminosities to increase as the Ca
H\&K break decreases. Fig. 4 and 5 show that this is indeed the case.
The Ca H\&K break value is significantly anti-correlated with both the
X-ray ($P > 99.9\%$) and radio core luminosity ($P > 99.9\%$). We used
the ASURV analysis package (Feigelson \& Nelson 1985) in the case of
censored data. The correlations remain in both cases very strong ($P >
99.9\%$) even if objects with Ca H\&K break values $C >0.4$ are
excluded, i.e., objects currently defined as radio galaxies. We also
verified by means of a partial correlation analysis that all our
significant luminosity - Ca H\&K break correlations were not induced
by a common redshift dependence. In the case of censored data we used
the algorithms developed by Akritas \& Siebert (1996). Note
additionally that, since we made it a requirement for our sources to
have information on both radio core and X-ray luminosity, Fig. 4 and 5
include the {\sl same} objects (83 sources).

We find a similarly strong ($P>99.9\%$) correlation between Ca H\&K
break value and X-ray luminosity if we consider BL Lacs and radio
galaxies from the DXRBS only (21 objects). Note that not all of these
sources have a measured radio core luminosity. Additionally,
Caccianiga et al. (1999) found a significant correlation between Ca
H\&K break value and X-ray luminosity for the BL Lacs and radio
galaxies from the REX survey, once sources residing in clusters of
galaxies were excluded. Since both the DXRBS and REX are homogeneuos
flux-limited samples, these results suggest that the luminosity - Ca
H\&K break correlations present in our sample are unlikely to be
caused by selection effects.

\subsection{Intrinsic luminosity range vs. orientation effects} 

\begin{figure}
\centerline{\psfig
{figure=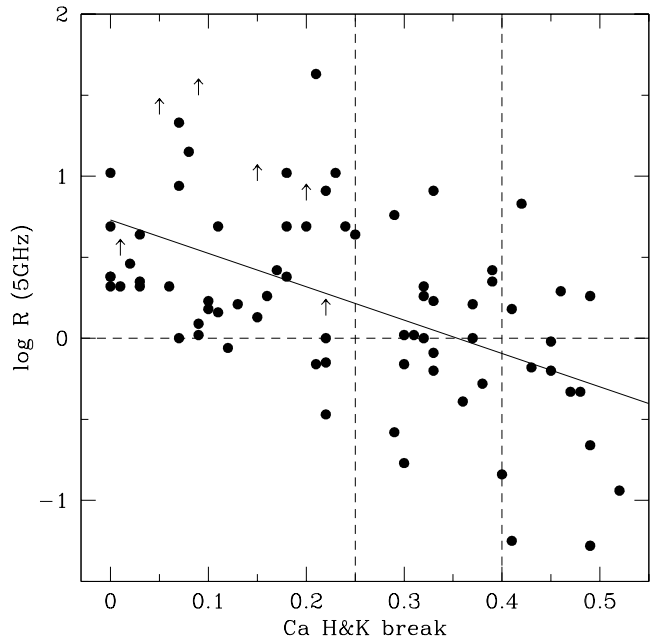,width=9cm}}
\caption{The core dominance parameter at 5 GHz vs. the Ca H\&K break value.
  Arrows denote lower limits. The solid line represents the observed
  correlation. The horizontal dashed line represents the locus of
  constant $R = L_{\rm rc}/L_{\rm ext} = 1$. The vertical dashed lines
  represent Ca H\&K break values of $C=0.25$ and 0.4 suggested by
  Stocke et al. (1991) and March\~a et al. (1996) respectively to
  discriminate between BL Lacs and radio galaxies.}
\end{figure}

So far we have shown by using observations in the optical, radio and
X-ray band that the decrease in Ca H\&K break is due to an increase in
the object's jet power. In the following we want to investigate {\sl
  why} BL Lac objects with a low Ca H\&K break have more powerful jets
than those with a higher Ca H\&K break value. There are two possible
explanations: 1. the spread in jet luminosities is intrinsic to the BL
Lac class, and/or 2. the spread in jet luminosities (and Ca H\&K break
value) is induced by orientation effects.

Which of these two possibilities causes the Ca H\&K break value to
correlate with jet luminosity can be best determined by using
information in the radio band. In this range, measurements on both
core and extended jet luminosities are available. This is important,
since the two powers are affected in a different way by orientation,
but similarly if the jet luminosity range is intrinsic. Beaming, i.e.,
a change in viewing angle, is known to cause only an increase in core
luminosity, but not to affect extended power, since the latter is
believed to be radiated isotropically. A change in intrinsic
luminosity, on the other hand, would be apparent as an increase in
both core and extended radio power (e.g., Giovannini et al. 1988) with
decreasing Ca H\&K break value. For the objects in our sample, no
significant ($P\sim46\%$) correlation is present between Ca H\&K break
value and extended radio power $L_{\rm ext}$. We note that, though we
have limited the extended radio powers of our objects to $L_{\rm ext}
< 10^{25.6}$ W/Hz, the distributions of both core and extended radio
powers have similar dispersions $\sigma = 0.7$ and 0.8 respectively.
Therefore, a priori there is no reason why the Ca H\&K break value
should correlate with radio core but not with extended radio emission.

On the other hand, if the Ca H\&K break value can indeed be related to
the object's viewing angle, it is expected to also correlate with the
radio core dominance parameter, thought to be a good indicator of
orientation. In Fig. 6 we plot for the objects in our sample the radio
core dominance parameter at 5 GHz vs. the Ca H\&K break value. The
former is defined as $R = L_{\rm rc}/L_{\rm ext}$. A significant ($P >
99.9\%$) linear anti-correlation, albeit with a large scatter, is
present between the two quantities. The correlation remains very
strong even if we eliminate objects with Ca H\&K break values $C >
0.4$.

Wolter et al. (2001) presented a plot similar to our Fig. 6 for BL
Lacs and FR I radio galaxies from the REX survey. They obtained for
their sample ($\sim 40$ objects) that Ca H\&K break value and radio
core dominance parameter (at 5 GHz) are only weakly anti-correlated.
However, they noted that their radio galaxies were more core-dominated
than classical \mbox{FR Is}. In fact, all their objects had radio core
dominance parameters $\log R \ge 0$. If we include in our analysis
only the core-dominated objects, we find that the correlation between
Ca H\&K break value and radio core dominance parameter also becomes
insignificant ($P=85\%$).

Fig. 6 shows further that the transition between core- and
lobe-dominated objects, i.e., between BL Lacs and FR I radio galaxies,
seems to occur rather smoothly as the Ca H\&K break value (and
therefore the viewing angle) changes, and we find that in the range
formerly defined as the BL Lac regime, i.e., $0 \le C \le 0.25$, only
11\% of the objects are lobe-dominated ($\log R < 0$), whereas this
ratio increases in the Ca H\&K break value ranges $0.25 < C \le 0.4$
and $C>0.4$ to 40\% and 69\% respectively. Therefore, the objects
newly included by March\~a et al. (1996) with Ca H\&K break values
$0.25 < C \le 0.4$ seem to represent the long-sought population
intermediate between the ``classical'' BL Lacs and the FR I radio
galaxies.

Further support for the assumption that the range in Ca H\&K break
values represents a range in viewing angles can be found by using the
results of Zirbel \& Baum (1995) and Hardcastle \& Worrall (1999).
Zirbel \& Baum (1995) obtained for a sample of 81 FR I radio galaxies
a mean radio core luminosity of $23.4 \pm 1.0$ W/Hz. This is
consistent with the value of $\log L_{\rm rc}= 23.8 \pm 0.3$ W/Hz that
the correlation in Fig. 5 gives for $C=0.5$, i.e., for the radio
galaxies.  Hardcastle \& Worrall (1999) could detect in ROSAT pointed
observations a core in the X-ray band for 15 of the FR I radio
galaxies studied (see their Table 4). For these we obtain a mean ROSAT
X-ray core luminosity $\log L_{\rm x} = 17.2 \pm 0.9$ W/Hz. This is in
good agreement with the value of $\log L_{\rm x} = 17.6 \pm 0.5$ W/Hz
that the correlation in Fig. 4 gives for $C=0.5$. Therefore, our
luminosity - Ca H\&K break correlations which we obtained including
both BL Lacs and FR I radio galaxies seem to reproduce well the
average radio and X-ray luminosities of (rather large samples of) FR I
radio galaxies.

\section{The relation between Ca H\&K break value and viewing angle}

\subsection{Luminosity ratios and viewing angle}

We now want to investigate how the Ca H\&K break value might be used
to constrain the viewing angle. We have then simulated for the radio
and X-ray band the expected luminosity ratio between the BL Lac object
and the parent radio galaxy as the viewing angle $\phi_{\rm GAL}$
changes. We used the formula given in Capetti \& Celotti (1999):
\begin{center}
$L_{\rm BL}/L_{\rm GAL} = [(1 - \beta \cdot {\rm cos} \phi_{\rm GAL})/
(1 - \beta \cdot {\rm cos} \phi_{\rm BL})]^{(p + \alpha)}$,
\end{center}
where $\beta$ is the bulk velocity in units of the speed of light, and
$\phi$ is the angle between the velocity vector and the line of sight.
We chose the case of a continuous jet ($p = 2$). For the simulations
in the radio band we assumed a radio spectral index $\alpha_{\rm
r}=0.2$, corresponding to the mean value for the objects in our
sample, while for the X-ray band we chose $\alpha_{\rm x}=1.4$,
following the results of Padovani \& Giommi (1996). The simulations
were performed assuming the three cases $\phi_{\rm BL} = 0^{\circ}$,
$10^{\circ}$, and $20^{\circ}$, and a changing $\phi_{\rm GAL}$ up to
a maximum value of $90^{\circ}$. In our case, the quantity $\phi_{\rm
BL}$ is fixed and represents the angle below which the jet/galaxy
ratio is so high that the Ca H\&K break value is constant and equal to
zero. Only $\phi_{\rm GAL}$ is assumed to change as the Ca H\&K break
value increases.  For each case of $\phi_{\rm BL}$ we assumed jet
Lorentz factors $\Gamma = 1/\sqrt{1 - \beta^2} = 2,3$ and 4. Our
results for the radio and X-ray band are shown in Fig. 7,
representatively for the case of $\phi_{\rm BL} = 10^{\circ}$ and
Lorentz factor $\Gamma = 3$.

\begin{table}
\begin{center}
{\bf Table 2.} Approximate average viewing angles of FR I galaxies
\\[0.1cm] \nobreak
\begin{tabular}{lccc} 
\hline 
\multicolumn{1}{c}{BL Lac angle:} & $0^{\circ}$ & $10^{\circ}$ & $20^{\circ}$ 
\\
\hline
$\Gamma=4$ & $32^{\circ}$ & $40^{\circ}$ & $58^{\circ}$ \\
$\Gamma=3$ & $44^{\circ}$ & $50^{\circ}$ & $66^{\circ}$ \\
$\Gamma=2$ & $72^{\circ}$ & $77^{\circ}$ & $>90^{\circ}$ \\ 
\hline
\end{tabular}
\end{center}
\end{table}

\begin{figure}
\centerline{\psfig
{figure=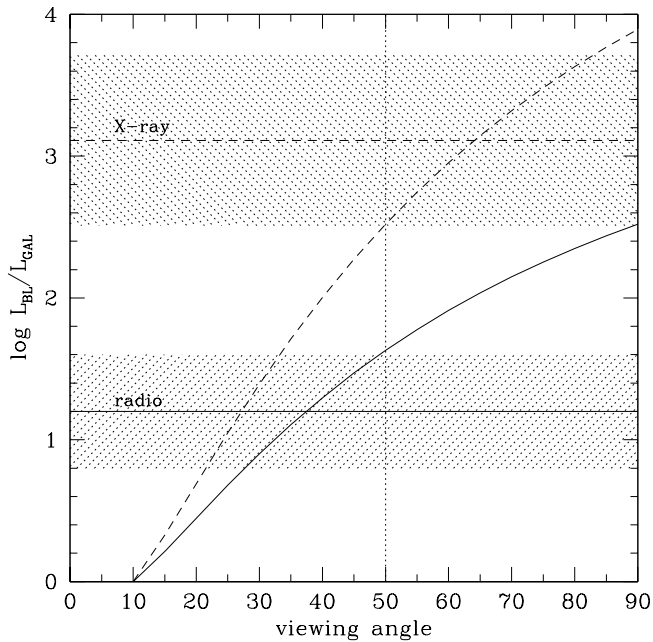,width=9cm}}
\caption{Simulated $L_{\rm BL}/L_{\rm GAL}$ ratios in the X-ray
  (dashed line) and radio band (solid line) for a Lorentz factor
  $\Gamma =3$, assuming a fixed BL Lac viewing angle $\phi_{\rm BL} =
  10^{\circ}$ and a changing FR I galaxy viewing angle $\phi_{\rm
    GAL}$ (see text for details). Horizontal lines indicate the ratios
  inferred from Fig. 4 and 5 in the X-ray (dashed line) and radio band
  (solid line) respectively. The $2\sigma$ error on the ratios is
  shown by the shaded areas. The vertical dotted line represents the
  FR I galaxy viewing angle at which the observed $L_{\rm BL}/L_{\rm
    GAL}$ ratios in the radio and X-ray band are reproduced
  approximately simultaneously.}
\end{figure}

From the best fits in Fig. 4 and 5 we get a maximum ratio of $\log
L_{\rm BL}/L_{\rm GAL} = 3.11 \pm 0.30$ and $1.20 \pm 0.20$ for the
X-ray and radio band respectively, where $L_{\rm BL}$ is the
luminosity at the Ca H\&K break value $C = 0$ and $L_{\rm GAL}$ the
luminosity at $C = 0.5$. Our simulations show that in all three cases
assumed for $\phi_{\rm BL}$ and $\Gamma$ it is possible to reproduce
these two ratios within their $2\sigma$ errors simultaneously, i.e.,
with the same Lorentz factor and approximately with the same maximum
viewing angle. Our results are listed in Table 2. The derived viewing
angles represent the angles under which we expect to observe on
average an FR I radio galaxy with a Ca H\&K break value of $C=0.5$.
These are not to be confused with maximum viewing angles for FR I
radio galaxies, which can be larger, or with the critical viewing
angle, i.e., the angle separating BL Lacs from FR I radio galaxies.
From Table 2 we see that the case of a starting BL Lac viewing angle
$\phi_{\rm BL} = 20^{\circ}$ and a Lorentz factor $\Gamma=2$ is
unphysical, since it gives typical viewing angles for radio galaxies
above $90^{\circ}$. Higher Lorentz factors than the ones simulated
here would result in smaller angles for the radio galaxies. Similarly
for the case of $p = 3$, describing a moving, isotropic source.

Our resulting viewing angles and Lorentz factors for FR I radio
galaxies agree with values obtained using independent methods,
validating our use of the Ca H\&K break value as an orientation
indicator. Giovannini et al. (2001) constrained the Lorentz factor for
parsec-scale jets in both low- and high-power radio galaxies to
$\Gamma = 3 - 10$ based on the scatter of their correlation between
core and total radio power. Urry \& Padovani (1995) obtained from
luminosity function studies of BL Lacs and FR I radio galaxies average
Lorentz factors of $\Gamma = 3 - 7$ and critical angles $\phi_{\rm c}
= 12 - 30^{\circ}$. The latter transform to average viewing angles for
the \mbox{FR Is} in the range $58 - 63^{\circ}$ if we use their
equation (A12). Chiaberge et al. (2000) derived from the observed
differences in core power between BL Lacs and \mbox{FR Is} with
similar extended radio powers average Lorentz factors $\Gamma \sim 5$,
if maximum viewing angles $\phi = 1/\Gamma$ and $60^{\circ}$ are
assumed for BL Lacs and \mbox{FR Is} respectively. Verdoes Kleijn et
al. (2002) used the correlation between radio core and H$\alpha+$[NII]
core emission for their sample of FR I radio galaxies and constrained
the average Lorentz factors to $\Gamma \sim 2 - 5$, assuming viewing
angles in the range $\phi=[30^{\circ},90^{\circ}]$.

\begin{figure}
\centerline{\psfig
{figure=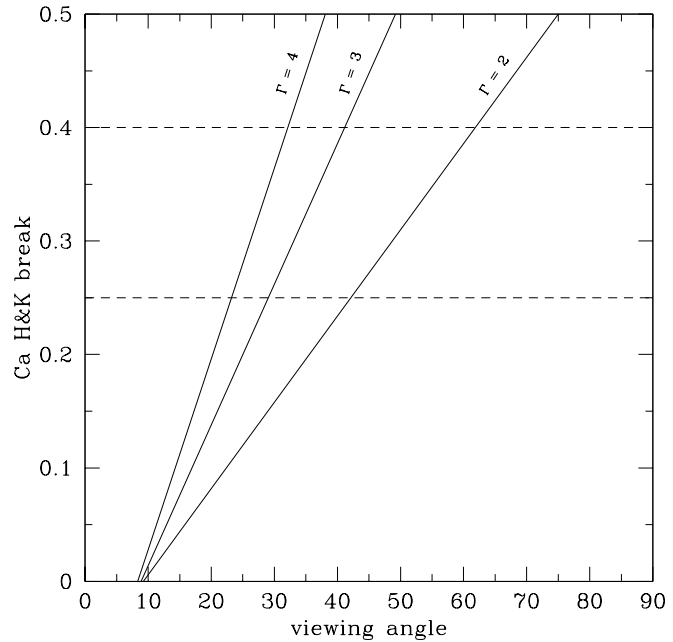,width=9cm}}
\caption{The correlation Ca H\&K break value vs. viewing angle for a
  starting viewing angle $\phi_{\rm BL}= 10^\circ$ and Lorentz factors
  $\Gamma = 2,3$ and 4 as obtained from connecting the correlations in
  Fig. 4 and 5 with our simulations exemplified in Fig. 7.}
\end{figure}

\subsection{Ca H\&K break value and viewing angle}

We want now to investigate how the Ca H\&K break value can be
converted to viewing angle. For this purpose, we determine from the
correlations in Fig. 4 and 5 for a range of Ca H\&K break values the
ratio $L_{\rm BL}/L_{\rm C}$ in the X-ray and radio band. We assume
$L_{\rm BL}$ to be the luminosity at the Ca H\&K break value $C = 0$,
and define $L_{\rm C}$ as the luminosity at $C = 0.1, 0.2, 0.3, 0.4$
and 0.5. We then convert the $L_{\rm BL}/L_{\rm C}$ ratios observed in
the two bands to a viewing angle using our simulations exemplified in
Fig. 7. Since each $L_{\rm BL}/L_{\rm C}$ ratio pair also corresponds
to an individual Ca H\&K break value, we obtain in this way a
correlation between Ca H\&K break and viewing angle. The linear fits
to the data points are shown in Fig. 8 for a jet of Lorentz factor
$\Gamma = 2, 3$ and 4, and assuming for $C=0$ a starting viewing angle
of $\phi_{\rm BL} = 10^{\circ}$.

From Fig. 8, we see that by extending the ``classical'' Ca H\&K break
value of 0.25 to 0.4 as proposed by March\~a et al. (1996), BL Lac
surveys include sources seen at larger viewing angles, and therefore
less beamed. This is supported by the recent results of Dennett-Thorpe
\& March\~a (2000), who find that BL Lac objects with $C>0.25$, the
so-called BL Lac candidates, are significantly less polarized at 8.4
GHz than BL Lacs with $C<0.25$.

\section{Low- and high-energy peaked BL Lacs}

The BL Lac class is currently divided into two subclasses: low- (LBL)
and high-energy peaked BL Lacs (HBL), i.e., objects with a synchrotron
emission peak located in the IR/optical and UV/soft X-ray band
respectively. This division was introduced by Padovani \& Giommi
(1995) after it was first discovered that BL Lacs detected in radio
(RBL) and X-ray surveys (XBL) had different radio-to-X-ray flux
ratios, and that this was due to the fact that their spectral energy
distributions (SED) had different shapes (Giommi, Ansari \& Micol
1995).

As of today the question of how the two BL Lac subclasses are
connected with each other regarding, e.g., their physical properties
or evolutionary behaviour, is a matter of fervent debate. A finding
that shaped strongly our perception of BL Lacs was made by Maraschi et
al. (1986). These authors found that radio- and X-ray-selected blazars
differed considerably in their radio luminosities, but had similar
X-ray powers. Based on this, they concluded that the X-ray radiation
was less beamed than the radio one, and that HBL being radio-weak were
objects viewed at larger angles than LBL. However, Sambruna, Maraschi
\& Urry (1996) later showed that the typical SED of an XBL cannot be
obtained from the SED of an RBL simply by changing the viewing angle
alone.

\begin{figure}
\centerline{\psfig
{figure=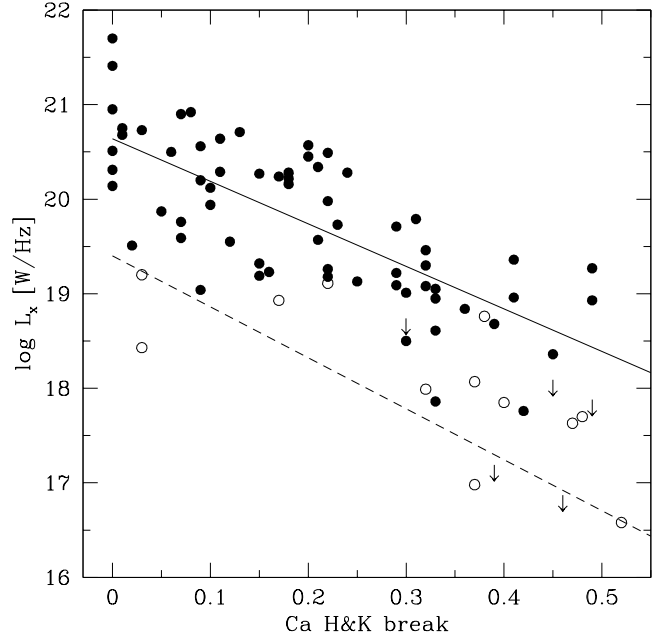,width=9cm}}
\caption{The 1 keV X-ray luminosity vs. the Ca H\&K break value. Open
  and filled circles denote LBL ($\log L_{\rm rc}/L_{\rm x} \ge 6$)
  and HBL ($\log L_{\rm rc}/L_{\rm x} < 6$) respectively. Arrows
  indicate LBL with upper limits on the X-ray luminosity. Solid and
  dashed lines represent the observed correlations for HBL and LBL
  respectively. Object 1241$+$735 is off the plot.}
\end{figure}

\begin{figure}
\centerline{\psfig
{figure=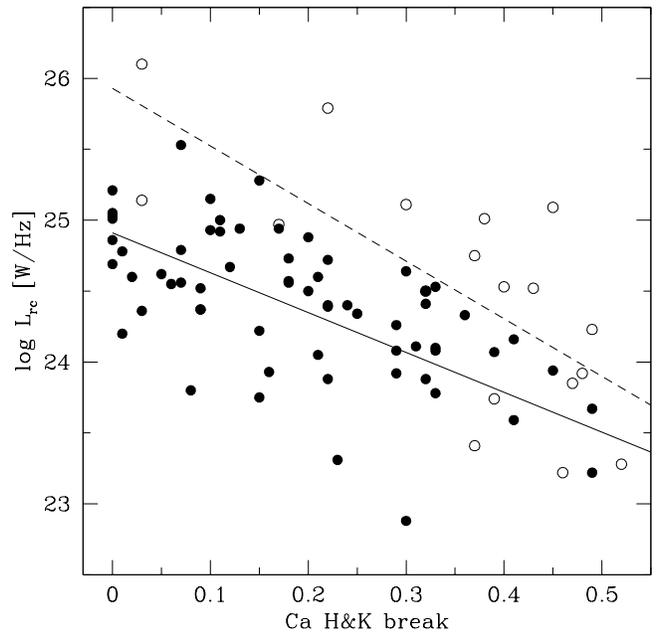,width=9cm}}
\caption{The radio core luminosity at 5 GHz vs. the Ca H\&K break
  value. Open and filled circles denote LBL ($\log L_{\rm rc}/L_{\rm
    x} \ge 6$) and HBL ($\log L_{\rm rc}/L_{\rm x} < 6$) respectively.
  Solid and dashed lines represent the observed correlations for HBL
  and LBL respectively.}
\end{figure}

If we divide the objects in our sample into LBL (defined by $\log
L_{\rm rc}/L_{\rm x} \ge 6$ [Padovani \& Giommi 1996]; 18 objects) and
HBL (defined by $\log L_{\rm rc}/L_{\rm x} < 6$; 65 objects), we find
similarly to the entire sample that their X-ray and radio core
luminosities are strongly anti-correlated with Ca H\&K break value
(Fig. 9 and 10 respectively). Note that we subdivided all the objects
in our sample, i.e., independent of Ca H\&K break value. We think that
this approach is justified since we are using the radio core instead
of the total radio power to define LBL and HBL. The significance level
is $P > 99.9\%$ for all four correlations. This finding allows us, on
one hand, to derive typical viewing angles for LBL and HBL, and, on
the other hand, to compare the luminosities of the two BL Lac
subclasses at different orientations.

\subsection{Viewing angles}

We performed the simulations described in Sect. 4.1 individually for
the two BL Lac subclasses.

The LBL and HBL in our sample have mean radio spectral indices
$\alpha_{\rm r}=0.1 \pm 0.1$ and $0.3 \pm 0.1$ respectively, not
significantly different ($P=93.7\%$) according to a Student's t-test.
Therefore, we assumed for the simulations in the radio band for both
LBL and HBL a radio spectral index $\alpha_{\rm r}=0.2$. On the other
hand, the X-ray spectral index is known to be different for the two
types of BL Lacs. Therefore, we performed the simulations in the X-ray
band independently for LBL and HBL and used, following the results of
Padovani \& Giommi (1996), $\alpha_{\rm x}=1.1$ and 1.5 respectively.

\begin{figure}
\centerline{\psfig
{figure=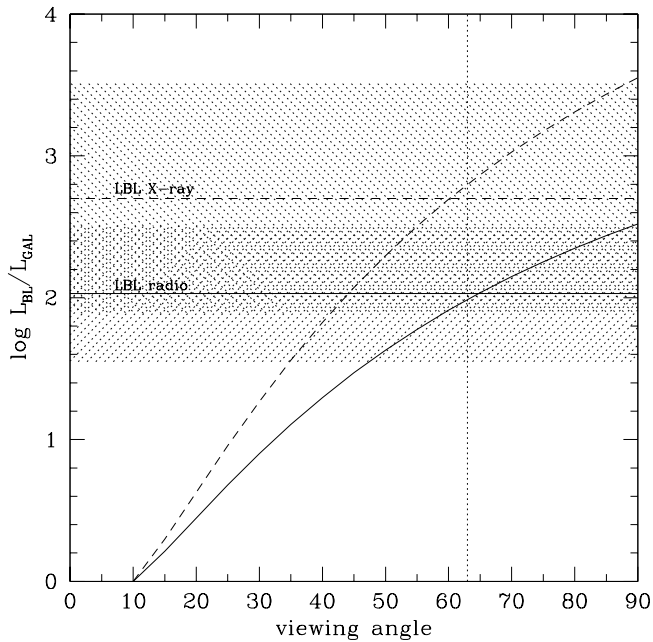,width=9cm}}
\caption{Simulated $L_{\rm BL}/L_{\rm GAL}$ ratios for LBL in the
  X-ray (dashed line) and radio band (solid line) for a changing
  viewing angle and a Lorentz factor $\Gamma =3$. Horizontal lines
  indicate the ratios inferred from Fig. 9 and 10 in the X-ray (dashed
  line) and radio band (solid line) respectively. The $1\sigma$ error
  on the ratios is shown by the shaded areas. The vertical dotted line
  represents the viewing angle at which the observed $L_{\rm
    BL}/L_{\rm GAL}$ ratios for LBL in the radio and X-ray band are
  reproduced approximately simultaneously.}
\end{figure}

\begin{figure}
\centerline{\psfig
{figure=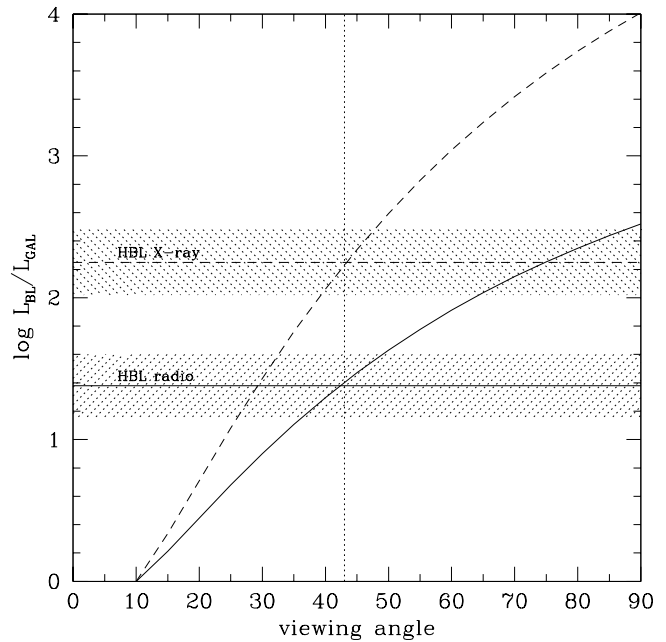,width=9cm}}
\caption{Simulated $L_{\rm BL}/L_{\rm GAL}$ ratios for HBL in the
  X-ray (dashed line) and radio band (solid line) for a changing
  viewing angle and a Lorentz factor $\Gamma =3$. Horizontal lines
  indicate the ratios inferred from Fig. 9 and 10 in the X-ray (dashed
  line) and radio band (solid line) respectively. The $1\sigma$ error
  on the ratios is shown by the shaded areas. The vertical dotted line
  represents the viewing angle at which the observed $L_{\rm
    BL}/L_{\rm GAL}$ ratios for HBL in the radio and X-ray band are
  reproduced approximately simultaneously.}
\end{figure}

From the best fits in Fig. 9 and 10 we get for LBL a maximum ratio of
$\log L_{\rm BL}/L_{\rm GAL} = 2.70 \pm 0.81$ and $2.03 \pm 0.48$ for
the X-ray and radio band respectively, where $L_{\rm BL}$ is the
luminosity at the Ca H\&K break value $C = 0$ and $L_{\rm GAL}$ the
luminosity at $C = 0.5$. For HBL, we obtain from the best fits in Fig.
9 and 10 ratios $\log L_{\rm BL}/L_{\rm GAL} = 2.25 \pm 0.26$ and
$1.38 \pm 0.22$ for the X-ray and radio band respectively.

Our simulations show that in all cases assumed for $\phi_{\rm BL}$ and
$\Gamma$ (see Sect. 4.1) it is possible to reproduce the ratios in the
radio and X-ray band for both LBL (Fig. 11) and HBL (Fig. 12)
simultaneously, i.e., with the same Lorentz factor and the same
maximum viewing angle. The viewing angles that we obtain in this way
are similar to the ones listed in Table 2. We note that in the case of
LBL we get somewhat higher viewing angles than for HBL. However, this
difference is not significant ($P=77.0\%$). A significantly larger
range in viewing angles would imply that LBL were more beamed than
HBL. This becomes clear if one recalls that in our simulations we
assumed the same Lorentz factor for LBL and HBL. Alternatively, we
could have fixed the range in viewing angles, which would have
resulted in larger Lorentz factors for LBL.

\subsection{Radio and X-ray luminosity differences}

\begin{table}
{\bf Table 3.} Radio core and X-ray luminosities$^{\star}$ 
of LBL and HBL derived from correlations in Fig. 9 and 10 \\
\begin{center}
\begin{tabular}{llccc} 
\hline 
& & LBL & HBL & $P$ \\ 
\hline
$C=0$   & $\log L_{\rm x}$ & $19.40 \pm 0.59$ & $20.65 \pm 0.12$ & $96.4\%$ \\
        & $\log L_{\rm rc}$ & $25.93 \pm 0.36$ & $24.90 \pm 0.11$ & $98.8\%$ \\
$C=0.5$ & $\log L_{\rm x}$ & $16.71 \pm 1.00$ & $18.39 \pm 0.28$ & $89.0\%$ \\
        & $\log L_{\rm rc}$ & $23.90 \pm 0.60$ & $23.51 \pm 0.25$ & $45.1\%$ \\
\hline
\end{tabular}
\end{center}
$^\star$ luminosities in units of [W/Hz]
\end{table}

\begin{table*}
{\bf Table 4.} Mean radio core and X-ray luminosities$^{\star}$ of 
LBL and HBL \\
\begin{minipage}{130mm}
\begin{tabular}{lccrcrcr} 
\hline 
& FR I HBL & FR I LBL & \multicolumn{1}{c}{$P$} & FR I\&II LBL & 
\multicolumn{1}{c}{$P$} & FR II LBL & \multicolumn{1}{c}{$P$} \\  
\hline
$\log L_{\rm x}$  & $19.77 \pm 0.10$ & $17.49 \pm 0.34$ & $>99.9\%$ 
                                     & $18.04 \pm 0.36$ & $>99.9\%$ 
                                     & $19.71 \pm 0.50$ & $8.0\%$ \\
$\log L_{\rm rc}$ & $24.36 \pm 0.08$ & $24.51 \pm 0.19$ & $57.9\%$ 
                                     & $25.14 \pm 0.26$ & $99.3\%$ 
                                     & $26.77 \pm 0.25$ & $>99.9\%$ \\
$z$               & $@0.21 \pm 0.02$ & $@0.10 \pm 0.02$ & $>99.9\%$   
                                     & $@0.23 \pm 0.05$ & $37.3\%$    
                                     & $@0.57 \pm 0.07$ & $>99.9\%$ \\
\hline
\end{tabular}
$^\star$ luminosities in units of [W/Hz]
\end{minipage}
\end{table*}

Our use of the Ca H\&K break value allows us to evaluate and compare
for the first time the radio and X-ray luminosities of LBL and HBL at
different orientations. In Table 3 we list the X-ray and radio core
powers resulting from the correlations in Fig. 9 and 10 for LBL and
HBL at small ($C=0$) and large ($C=0.5$) Ca H\&K break values and
therefore viewing angles. From this we get that at large viewing
angles LBL and HBL have similar radio core and X-ray powers, i.e.,
they reside in FR I radio galaxies with similar properties (radio core
and X-ray luminosities), while at relatively small viewing angles the
two BL Lac subclasses differ significantly in their luminosities. In
this case we get that LBL are $\approx 10$ times more luminous in the
radio and by a similar factor less luminous in the X-ray band than
HBL. We stress that the objects used in this work form an
heterogeneous sample of sources from different surveys with widely
different flux limits. Therefore, although we believe that these
luminosity differences are present, their precise values might depend
on the selected objects. In the following we want to expand on the
influence of selection effects on the luminosity differences between
LBL and HBL.

For this purpose we have also included BL Lacs with extended radio
powers more typical of FR II radio galaxies ($L_{\rm ext} > 10^{25.6}$
W/Hz). Furthermore, we distinguished between the following three
cases: 1. a comparison between HBL (65 objects) and LBL (18 objects),
where both have extended radio emissions typical of FR I radio
galaxies; 2. a comparison between HBL (65 objects) and LBL (25
objects), where the latter are selected independent of extended radio
emission; 3. a comparison between HBL (65 objects) and LBL (7
objects), where the latter have extended radio emissions typical of FR
II radio galaxies. The most important difference between these three
cases is that we compare HBL and LBL that are first matched and then
not matched in extended radio power, i.e., have a similar parent
population.

For the three cases, we obtain the following results, illustrated in
Table 4: 

1. In the first case, where we compare LBL and HBL with similar
extended radio powers, we get that they have similar mean radio core
luminosities, while HBL have higher mean X-ray luminosities than
LBL. In this case, we get that LBL have a significantly lower mean
redshift than HBL.

2. In the second case, where we compare LBL and HBL with somewhat
different extended radio powers, we get that LBL have higher mean
radio core luminosities and lower mean X-ray luminosities than HBL. In
this case, LBL and HBL have similar mean redshifts.

3. In the third case, where we compare LBL with high extended radio
powers and HBL with low extended radio powers, we get that the two BL
Lac subclasses have similar mean X-ray luminosities, while LBL have
higher mean radio core luminosities than HBL. In this case, LBL have a
significantly higher mean redshift than HBL.

These comparisons show that the resulting luminosity differences
between LBL and HBL seem to depend strongly on the samples chosen. In
particular, our three cases illustrate that, the more LBL and HBL
differ in their extended radio powers, the less they differ in their
X-ray powers. We note that these luminosity differences cannot be due
only to a redshift effect. If that were the case, in fact,
higher-redshift samples would be more luminous in all bands, contrary
to what observed.

We can further illustrate these selection effects by using two
different samples: the sample of Maraschi et al. (1986) that contains
LBL and HBL with similar X-ray powers, and the complete sample of BL
Lacs from the DXRBS which is radio-flux limited and therefore contains
LBL and HBL with similar radio powers. Nearly half of the objects in
the sample used by Maraschi et al. (1986) are strong-lined objects,
i.e., radio quasars. Therefore, in order to compare their results with
ours we selected from their sample only the BL Lac objects. Similarly
to their results for radio- and X-ray-selected blazars, we get that
their LBL (17 objects) and HBL (11 objects) have similar mean X-ray
powers, but significantly different ($P>99.9\%$) mean (total) radio
luminosities $\log L_{\rm r}=26.61 \pm 0.25$ and $24.56 \pm 0.08$ W/Hz
respectively. The mean redshifts are $z=0.35 \pm 0.08$ and $0.12 \pm
0.03$ for LBL and HBL respectively, different at the 98.8\% level.
This result is similar to our case 3, indicating that these authors
have compared radio-strong LBL with radio-weak HBL.

For the sample of DXRBS BL Lacs, we get that LBL (20 objects) and HBL
(12 objects) have similar mean radio powers, but significantly
different ($P=98.7\%$) mean X-ray powers $\log L_{\rm x} = 18.98 \pm
0.23$ and $20.00 \pm 0.32$ W/Hz respectively. In this case LBL and HBL
have similar mean redshifts. This result shows that our first case can
be reproduced with a radio-flux limited sample.

Now the question arises: ``Which is the best approach to clarify what
are the intrinsic luminosity differences between LBL and HBL in a
given band?''. We think that this can be answered in a physically
meaningful way by comparing LBL and HBL with a similar parent
population, i.e., with similar extended radio powers. Ideally, if
information on the Ca H\&K break value is available, one should also
take into account orientation effects by separating sources according
to viewing angle.

\section{Discussion and Conclusions}
 
We have suggested that the Ca H\&K break value of BL Lacs and FR I
radio galaxies is a direct indicator of viewing angle. We have based
this on the strong anti-correlation between Ca H\&K break value and
optical, radio, X-ray jet luminosity, and radio core dominance
parameter. We have excluded the possibility that the observed range in
radio jet power is intrinsic and not due to orientation, since we
could not find a similarly strong correlation between Ca H\&K break
value and extended radio emission for our sample of objects.

Stocke et al. (1991) introduced a limit of 25\% on the allowed Ca H\&K
break value for BL Lacs, which was later expanded by March\~a et al.
(1996) up to a value of 40\%, in connection with the strength of
observed emission lines. Our result implies that those BL Lacs with Ca
H\&K break values $0.25 \le C \le 0.4$, termed by March\~a et al.
(1996) ``BL Lac candidates'', represent in fact the long-sought
population of objects with viewing angles intermediate between the
``classical'' BL Lacs and FR I radio galaxies.

Which limit on the Ca H\&K break value should be chosen to separate BL
Lacs from FR I radio galaxies? In general, BL Lacs are those objects
viewed at angles smaller than a certain critical angle, which has been
defined in the literature as the angle for which the radio core
dominance parameter is equal to 1 (Urry \& Padovani 1995). From the
correlation between the radio core dominance parameter and Ca H\&K
break value illustrated in Fig. 6 we infer that a Ca H\&K break value
$\sim0.35$ would then be appropriate to separate BL Lacs
(core-dominated) from radio galaxies (lobe-dominated). We note that
this value is very close to $0.4$, proposed by March\~a et al. (1996).

We have shown that the Ca H\&K break value is a suitable indicator of
orientation. So far, only one other such indicator was known: the
radio core dominance parameter. However, the determination of this
quantity usually requires dedicated radio observations, which are time
consuming and not always available. Therefore, our result that we can
constrain the viewing angle of a BL Lac or FR I radio galaxy from such
a common astrophysical observation as its optical spectrum will be a
considerable advantage in our studies of unified schemes.

We have shown for our full sample that radio and X-ray jets of BL Lacs
and FR I radio galaxies have similar Lorentz factors and are viewed
under similar angles, namely the radio and X-ray Doppler factors are
the same (within the errors). This result becomes even more
significant if we separate our sample into low- (LBL) and high-energy
peaked BL Lacs (HBL).

We have also shown that LBL and HBL have jets with similar Lorentz
factors, viewed under similar angles, i.e., their Doppler factors are
similar. This result is in agreement with the ``different peak
energy'' scenario (Padovani \& Giommi 1995), which claims that the
only difference between LBL and HBL is the frequency position of their
synchrotron emission peak. However, we note that there is a hint in
our data that LBL might be more beamed than HBL, i.e., they might have
either larger Lorentz factors or span a larger range in viewing
angles. But given that our sample is heterogeneous, this might be
induced by selection effects.

Our finding that the Ca H\&K break value is directly related to
viewing angle has allowed us to compare for the first time the
luminosities of the two types of BL Lacs at different orientations. We
derive that FR I radio galaxies harbouring LBL and HBL have similar
radio core and X-ray luminosities. At small viewing angles, LBL have
radio cores $\approx 10$ times more powerful than HBL ones, while the
opposite is true in the X-ray band. These two results combined appear
to be at odds with our previous result that LBL and HBL have similar
Doppler factors. We attribute this apparent contradiction to small
number statistics. Note in fact that HBL-like and LBL-like \mbox{FR
  Is} also differ in their powers in the same sense as HBL and LBL but
their differences are not significant due to the larger errors at $C
\sim 0.5$.

We have also discussed the issue of selection effects on the study of
luminosity differences between HBL and LBL, and have shown that the
comparison to be meaningful has to be done for samples with similar
parent populations, i.e., similar extended radio powers.

To summarize, we have reached the following conclusions:

\begin{enumerate}

\item The value of the Ca H\&K break in BL Lacs and FR I radio
galaxies decreases with increasing jet power.

\item The increase in jet power is caused by a change in viewing
angle.

\item BL Lacs with Ca H\&K break values $0.25 \le C \le 0.4$, termed
by March\~a et al. (1996) ``BL Lac candidates'', are the long-sought
population with viewing angles intermediate between the ``classical''
BL Lacs and FR I radio galaxies.

\item BL Lacs and FR I radio galaxies have radio and X-ray jets with
similar Lorentz factors ($\sim 2-4$), and viewed under similar angles,
i.e., their jets have similar radio and X-ray Doppler factors.

\item The two types of BL Lacs, LBL and HBL, have jets with similar
Lorentz factors, which are viewed under similar angles, i.e., they are
affected by beaming in a similar way.

\item LBL and HBL reside in FR I radio galaxies with similar radio
core and X-ray powers. However, at small viewing angles, LBL have
radio cores $\approx 10$ times more powerful than HBL ones, while the
opposite is true in the X-ray band.

\end{enumerate}

In future work we plan to investigate the emission line properties of
BL Lacs in order to understand in what way they are related to their
more powerful siblings, the flat spectrum radio quasars. With this we
hope to be able to give a more complete picture of what a BL Lac
really is.

\section*{Acknowledgements}

We would like to thank Eric Perlman, Manfred Stickel, and Clive
Tadhunter for providing several spectra in electronic format.
H.L. acknowledges financial support from the Deutscher Akademischer
Austauschdienst (DAAD) and the STScI DDRF grant D0001.82260.

\end{document}